\documentclass[12pt]{article}

\newcommand{\be}{\begin{equation}}
\newcommand{\ee}{\end{equation}}

\newcommand{\ba}{\begin{eqnarray}}
\newcommand{\ea}{\end{eqnarray}}

\newcommand{\VEV}[1]{\left\langle {#1} \right\rangle}

\newcommand{\e}{\,{\rm e}}

\begin{document}
\addtolength{\baselineskip}{.7mm}
\thispagestyle{empty}

\begin{flushright}
STUPP--02--168 \\ September, 2002
\end{flushright}
\vspace{.5cm}

\begin{center}
{\large{\bf{
Logarithmic Correlation Functions in Liouville Field Theory
}}}
\\[20mm]

{\sc Shun-ichi Yamaguchi}
\footnote{\ \tt 
E-mail: syama@post.saitama-u.ac.jp
} \\[10mm]

{\it Department of Physics, Faculty of Science, Saitama University
\\[2mm]
Saitama 338-8570, Japan} \\[4mm]

and \\[4mm]

{\it Department of Physics, School of Science, Kitasato University
\\[2mm]
Sagamihara 228-8555, Japan} \\[28mm]

{\bf Abstract} \\[7mm]
{\parbox{14cm}{\hspace{5mm}
We study four-point correlation functions with logarithmic behaviour 
in Liouville field theory on a sphere, which 
consist of one kind of the local operators. 
We study them as non-integrated correlation functions 
of the gravitational sector of two-dimensional quantum gravity 
coupled to an ordinary conformal field theory 
in the conformal gauge. 
We also examine, in the $(p,q)$ minimal conformal field theories, 
a condition of the appearance of logarithmic correlation functions 
of gravitationally dressed operators. 
}} \\[5mm]

\end{center}




\newpage
\setcounter{equation}{0}

One of physical examples 
in logarithmic conformal field theories \cite{LCFT1}--\cite{LCFT3} is 
the gravitationally dressed conformal field theory. 
Two-dimensional quantum gravity coupled to 
a free massless Majorana fermion field theory 
in the light-cone gauge had been studied in Ref.\ \cite{BK}, 
where the non-integrated four-point correlation function 
of the gravitationally dressed operators, which has 
logarithmic behaviour, was obtained. 
The origin of the logarithms of such models is 
Liouville field theory which describes 
the gravitational sector of the system. 
However, correlation functions with logarithmic behaviour 
in the Liouville field theory are not sufficiently understood so far. 

The purpose of this letter is to study 
four-point correlation functions with logarithmic behaviour 
in Liouville field theory on a sphere. 
We consider a certain class of four-point correlation functions 
of local Liouville operators, which can be regarded as 
non-integrated correlation functions of the gravitational sector 
of two-dimensional quantum gravity coupled to 
an ordinary conformal field theory in the conformal gauge. 
We obtain the correlation functions with logarithmic behaviour 
and find the appearance of logarithms is the same mechanism as that 
in the Coulomb-gas construction of the correlation function 
in the $c=-2$ conformal field theory \cite{HY}. 
We also examine a condition of the appearance 
of logarithmic correlation functions 
consist of one kind of gravitationally dressed operators 
in the $(p,q)$ minimal conformal field theories, 
and obtain the correlation function. 
The appearance of the correlation function 
with logarithmic behaviour in the gravitational dressing 
of the Majorana fermion field theory \cite{BK} can be also understood 
in our formulation. 

We first consider two-dimensional quantum gravity coupled to 
a conformal field theory on a sphere. 
The conformal field theory can be regarded as 
the matter part in the system. 
After conformal gauge fixing, the gravitational sector is 
described by the Liouville field theory with the action \cite{DDK}
\be
S_{\rm L}[\hat g, \phi] = 
{1 \over 8\pi} \int d^2 \xi \sqrt{\hat g} \left(
\hat g^{\alpha \beta} \partial_\alpha \phi \partial_\beta \phi 
- Q \hat R \phi + 4 \mu \e^{\alpha \phi} \right), 
\label{action}
\ee
where $\hat R$ is the scalar curvature 
with a fixed reference metric $\hat g_{\alpha \beta}$ and 
$\mu$ is the renormalized cosmological constant. 
If the two parameters $Q$ and $\alpha$ satisfy the condition 
\be
Q = - \alpha - {2 \over \alpha} \,, 
\label{CFTC}
\ee
the Liouville field theory (\ref{action}) describes 
a conformal field theory with the central charge \cite{CT}
\be
c_{\rm L} = 1 + 3 Q^2. 
\ee
The primary field in the theory takes the form $\e^{\beta \phi}$ 
and which has the conformal weight \cite{CT},\cite{SP} 
\be
h_\beta = -{1 \over 2} \beta^2 -{1 \over 2} \beta Q \,. 
\label{primary}
\ee
The two Eqs.\ (\ref{CFTC}) and (\ref{primary}) imply that 
the conformal weight $h_\alpha$ of the operator $\e^{\alpha \phi}$ 
in the action (\ref{action}) is one. 

When we consider the case that the matter part is realized by 
the $(p,q)$ minimal conformal field theory which has 
the central charge \cite{BPZ}
\be
c_{p,q} = 1 - 6{(p-q)^2 \over pq} \,, 
\label{pqc}
\ee
the parameters $Q$ and $\alpha$ are respectively given by \cite{DDK} 
\be
Q = -\alpha_+ -\alpha_- \,, 
\ee
where 
\be
\alpha_+ = \alpha = -\sqrt{2q \over p} \,, \quad 
\alpha_- = {2 \over \alpha_+} = - \sqrt{2p \over q} \,. 
\label{alphapm}
\ee
For each primary field $\Phi_{r,t}(\xi)$ 
with the conformal weight \cite{BPZ} 
\be
\Delta_{r,t} = {1 \over 8} \left[(r \alpha_- -t \alpha_+)^2
- (\alpha_- -\alpha_+)^2 \right] 
\qquad (\, t \alpha_+ > r \alpha_- \,) \,, 
\ee
there exists a gravitationally dressed operator of the form \cite{DDK} 
\be
O_{r,t}(\xi) = \e^{\beta \phi(\xi)} \Phi_{r,t}(\xi) \,. 
\label{physop}
\ee
The parameter $\beta$ is fixed by the requirement that 
the operator $O_{r,t}(\xi)$ is a primary field 
with the conformal weight one \cite{DDK}: 
\be
\beta =\beta_{r,t} ={1 \over 2}(1-r) \alpha_- 
+ {1 \over 2}(1+t) \alpha_+ \,. 
\label{conditionbeta}
\ee
As discussed in Ref.\ \cite{SP}, in general, 
only when the parameter $\beta$ satisfies 
the condition ${\rm Re} \,\beta > -{1 \over 2}Q$, 
the operator $\e^{\beta \phi(\xi)}$ can be interpreted as 
a local operator. 
All the operators (\ref{physop}) have real $\beta_{r,t}$ with 
$\beta_{r,t} > -{1 \over 2}Q$. Therefore they are local operators. 
The cosmological term operator $\e^{\alpha \phi(\xi)}$ 
in the action (\ref{action}) is a particular case of 
the operator (\ref{physop}) with $\Phi_{1,1}(\xi)$. 

We now consider non-integrated four-point correlation functions 
of the gravitationally dressed operators 
on a sphere with fixed area $A$. 
In order to study general properties of the correlation functions, 
we use local operators $\e^{\beta_i \phi(\xi)}$ with 
real $\beta_i$ which is not fixed by Eq.\ (\ref{conditionbeta}). 
On the complex plane, the correlation functions are separated into 
the Liouville part and the matter part as 
\be
\VEV{\prod_{i=1}^4 O_i (z_i, \overline z_i)}_A 
= \prod_{i=1}^4 \VEV{\e^{\beta_i \phi(z_i, \overline z_i)}}_{\! A}
\VEV{\Phi_i (z_i, \overline z_i)}. 
\label{total}
\ee
The Liouville expectation value is given by \cite{DDK} 
\be
\VEV{ \prod_{i=1}^4 
\e^{\beta_i \phi(z_i, \overline z_i)} }_{\! A}
= \int {\cal D} \phi \e^{-S_{\rm L} [\hat g, \phi]} 
\prod_{i=1}^4 \e^{\beta_i \phi(z_i, \overline z_i)}
\, \delta \left( \int d^2 u \sqrt{\hat g} 
\e^{\alpha \phi (u, \overline u)} - A \right). 
\ee
After integrating 
the zero mode $\phi_0$ $(\phi = \phi_0 + \tilde \phi)$, 
we obtain 
\be
\VEV{ \prod_{i=1}^4 
\e^{\beta_i \phi(z_i, \overline z_i)} }_{\! A}
= {1 \over |\alpha|} A^{-s-1}\,\tilde G_{\rm L}^{(s)}, 
\ee
where 
\be
\tilde G_{\rm L}^{(s)} 
= \VEV{
\prod_{i=1}^4 \e^{\beta_i \tilde \phi(z_i, \overline z_i)} 
\left(\int d^2 u \e^{\alpha \tilde\phi(u, \overline u)}\right)^{\! s}
}. 
\label{afterzero}
\ee
The parameter $s$ is 
\be
s = -{1 \over \alpha} \left( Q + \sum_{i=1}^4 \beta_{i} \right)
\label{s}
\ee
and the expectation value (\ref{afterzero}) of 
the non-zero modes $\tilde\phi$ is defined by using 
the free action 
\be
S_0 [\tilde \phi] = 
{1 \over 8 \pi} \int d^2 z \,
\partial_\alpha \tilde \phi \, \partial^\alpha \tilde \phi \,. 
\ee
In general, the parameter $s$ takes a generic real value. 
When $s$ is a non-negative integer, 
we can evaluate the expectation values (\ref{afterzero}). 
The correlation function (\ref{total}) with a 
generic real $s$ can be obtained by an analytic continuation 
in $s$ from that obtained for non-negative integers \cite{s}. 
However, the analytic continuation is possible only 
after performing the integrals of positions 
of the operators $O_i (z_i, \overline z_i)$ in (\ref{total}). 
Our purpose is to study properties 
of the non-integrated correlation functions, 
so in the following, we restrict ourselves to the cases 
with $s \in {\bf Z}_+$ since the case with $s=0$ is trivial. 

The non-zero mode expectation value (\ref{afterzero}) can be 
evaluated by using a similar technique as is used in 
the Coulomb-gas construction \cite{DF} of 
the minimal conformal field theories. 
In this way, $\tilde G_{\rm L}^{(s)}$ can be represented as 
\ba
\tilde G_{\rm L}^{(s)}
& \!\!\! = & \!\!\! \prod_{1 \leq i < j \leq 4} 
|z_i - z_j|^{-2(h_i +h_j) + {2 \over 3} h} \, 
|x|^{2(h_1 + h_2)-{2 \over 3} h -2 \beta_1 \beta_2}
\nonumber \\
& \!\!\! & \!\!\! \times \, |1-x|^{2(h_2 + h_3) - {2 \over 3} h 
-2 \beta_2 \beta_3} \, 
I^{(s)}(-\alpha\beta_1, -\alpha\beta_3, -\alpha\beta_2; 
- \textstyle{1 \over 2} \alpha^2; x) \,, 
\ea
where\ $h = \sum_{i=1}^4 h_i$, 
$x ={(z_1 -z_2)(z_3 -z_4) \over (z_1 -z_3)(z_2 -z_4)}$ and 
\be
I^{(s)} (a, b, c; \rho; x) 
= \int \prod_{i=1}^s d^2 u_i \prod_{i=1}^s 
\left[ |u_i|^{2a} |1-u_i|^{2b} |u_i-x|^{2c} \right] 
\prod_{1 \leq i < j \leq s} |u_i - u_j|^{4\rho} \,. 
\ee
The integral representation $I^{(s)} (a, b, c; \rho; x)$ 
can be transformed into a sum of squares of line integrals 
\be
I^{(s)} (a, b, c; \rho; x) 
\,=\, 
\sum_{k=0}^s X^{(s)}_k 
\left| I_k^{(s)} (a, b, c; \rho; x) \right|^2, 
\ee
where 
\ba
I_k^{(s)} (a, b, c; \rho; x) 
& \!\!\! = & \!\!\! 
\int_0^x \prod_{i=1}^k du_i \int_1^\infty \prod_{j=k+1}^s du_j 
\prod_{i=1}^s u_i^a 
\prod_{1 \leq i < j \leq s} (u_i - u_j)^{2\rho}
\nonumber \\
& \!\!\! & \!\!\! \quad \times \prod_{i=1}^k (1-u_i)^b (x-u_i)^c 
\prod_{j=k+1}^s (u_j-1)^b (u_j-x)^c \,. 
\label{Iks}
\ea
The explicit forms of the coefficients $X_k^{(s)}$ can be 
found in Ref.\ \cite{DF}. 

To study concrete examples in detail 
and to compare our results with that of Ref.\ \cite{BK} later, 
from now on, 
we consider the correlation functions consist of 
one kind of Liouville operators, 
i.e.\ $\beta_1 =\beta_2 =\beta_3 =\beta_4 \equiv \beta$ 
in Eq.\ (\ref{afterzero}). 
Then, from Eq.\ (\ref{s}), the parameter $\beta$ takes the values 
\be
\beta = -{1 \over 4}(s-1) \alpha + {1 \over 2 \alpha} \,. 
\ee
Before evaluating the correlation functions explicitly, we examine 
the $x \rightarrow 0$ behaviours of $I_k^{(s)} (a,b,c;\rho;x)$. 
They can be obtained by changing the 
integration variables as $u_i = x w_i$ for $i=1, \cdots, k$ 
in Eq.\ (\ref{Iks}). 
For $s=2,3,4,\cdots$, we obtain 
\be
I_k^{(s)} (a, b, c; \rho; x) 
\, \sim \, x^{{1 \over 2} k(s-k) \alpha^2}. 
\ee
Therefore, 
\be
\tilde G_{\rm L}^{(s)}
\,\sim\,
\sum_{k=0}^s X^{(s)}_k |x|^{-2 \beta^2 + k(s-k) \alpha^2} 
\quad {\rm for}\ \ x \rightarrow 0 \,. 
\ee
The singularities obtained above are those of 
approaching the two operators $\e^{\beta \tilde \phi}$ and $k$ of 
$s$ operators $\int d^2u \e^{\alpha \tilde \phi}$ one another 
in the non-zero mode expectation value (\ref{afterzero}) since 
the exponents of $|x|$ can be expressed as 
\be
-2 \beta^2 + k(s-k) \alpha^2 
= -2h_\beta -2h_\beta +2h_{\beta + \beta +k \alpha} \,. 
\ee
On the other hand, we find that $I_k^{(1)} (a, b, c; 0; x)$, 
which is $I_k^{(s)} (a, b, c; \rho; x)$ with $s=1$, 
has no power singularities for $x \rightarrow 0$, 
i.e.\ a $x$-independent constant. 
This is in contrast to the case with $s =2,3,4,\cdots$, and 
suggests the existence of logarithmic singularities. 

Let us express the case with $s=1$ more explicitly. 
The concrete form of $\tilde G_{\rm L}^{(1)}$ can be obtained by 
fixing $z_1 =0$, $z_2 = x$, $z_3 =1$, $z_4 =\infty$, and is 
\be
\tilde G_{\rm L}^{(1)} 
= F_{\rm L} (x, \overline x) \prod_{1 \le i < j \le 4} 
|z_i -z_j|^{-{4 \over 3} h_\beta}
\ee
where 
\be
F_{\rm L} (x, \overline x) 
= |x(1-x)|^{- 2 \beta^2 +{4 \over 3} h_\beta}
\, I^{(1)} (-\alpha\beta, -\alpha\beta, -\alpha\beta; 0; x)
\ee
and $\beta ={1 \over 2 \alpha}$. 
Here, one encounters an indeterminate form 
\be
I^{(1)} (-\alpha\beta, -\alpha\beta, -\alpha\beta; 0; x) 
= {\textstyle I^{(1)} 
\left(-{1 \over2}, -{1 \over2}, -{1 \over2}; 0; x\right) }
\sim {0 \over 0}
\ee
since $I_0^{(1)} = I_1^{(1)}= \pi F \left(
{1 \over2}, {1 \over2}, 1; x \right)$ and 
$X_0^{(1)} = - X_1^{(1)} = -{1 \over \sin (- \pi)}$. 
This indeterminate form had been evaluated 
in Refs.\ \cite{Fl} and \cite{HY}. 
The result is 
\ba
& & \textstyle{ 
I^{(1)} \left(-{1 \over 2}, -{1 \over 2}, -{1 \over 2}; 0; x \right) 
}
\nonumber \\
& & \qquad 
= \textstyle{
-\pi \ln|x|^2 \left| {\textstyle 
F \left({1 \over 2},{1 \over 2},1;x \right)
} \right|^2 
-2 {\textstyle 
F \left({1 \over 2},{1 \over 2},1;x \right)
}\,H(\overline x)
-2 {\textstyle 
F \left({1 \over 2},{1 \over 2},1;\overline x \right)
} \, H(x) \,, \qquad \quad
}
\ea
where 
\be
H(x) 
= \sum_{n=0}^\infty \left[\Gamma(n+{1 \over 2}) \over n! \right]^2 
\left[\psi(n+ {\textstyle {1 \over 2}}) -\psi(n+1)\right] x^n. 
\ee
Thus we find that the appearance of 
the correlation functions with logarithmic behaviour 
in Liouville field theory is the same mechanism as that 
in the Coulomb-gas construction of the correlation 
function $\VEV{\mu (z_1)\,\mu (z_2)\,\mu (z_3)\,\mu (z_4)}$ in 
the $c=-2$ conformal field theory \cite{HY}. 
The explicit form of the correlation functions is 
\be
\VEV{ \prod_{i=1}^4 
\e^{\beta \phi(z_i, \overline z_i)} }_{\! A}
= {1 \over |\alpha| A^2} \,|x (1-x)|^{-{1 \over 2 \alpha^2}}
\, |(z_1 -z_3)(z_2 -z_4)|^{-{3 \over 2 \alpha^2} -1} \, I^{(1)} 
\textstyle{ 
\left(-{1 \over 2}, -{1 \over 2}, -{1 \over 2}; 0; x \right), 
}
\label{result}
\ee
where 
\be
\beta ={1 \over 2 \alpha} \,. 
\label{ourbeta}
\ee
We also find that the correlation functions (\ref{result}) exist 
in the theory with the central charge 
\be
c_{\rm L} = 13 + 3 \alpha^2 + {12 \over \alpha^2} \,. 
\ee
Note that the local operator condition $\beta > -{1 \over 2}Q$ 
for $\beta$ in Eq.\ (\ref{result}) is always satisfied since 
the cosmological term operator 
in the action (\ref{action}) has $\alpha$ satisfying 
the codition $\alpha > -{1 \over 2}Q$. 
Actually, the parameter $\alpha$ can takes continuum values 
in the region $- \sqrt{2} < \alpha <0$, 
and the logarithmic theory obtained above has 
the central charge $c_{\rm L} > 25$. 

Now, to apply our result to the correlation function 
in the gravitational dressing of 
free massless Majorana fermion field theory, 
we finally consider the type of the correlation function 
$\VEV{\prod_{i=1}^4 O_{r,t}(z_i, \overline z_i)}_A$. 
The matter part 
$\VEV{\prod_{i=1}^4 \Phi_{r,t} (z_i, \overline z_i)}$ 
does not vanish since, for such correlation functions, 
the charge neutrality condition 
in the Coulomb-gas construction can be satisfied by inserting 
screening charges $(Q_+)^{t-1}$ and $(Q_-)^{r-1}$ \cite{DF}. 
Corresponding Liouville operators 
are $\e^{\beta_{r,t} \phi(z_i, \overline z_i)}$, 
in which the value of $\beta_{r,t}$ is 
fixed by Eq.\ (\ref{conditionbeta}). 
From Eqs.\ (\ref{conditionbeta}) and (\ref{ourbeta}), we obtain 
a relation among the integer parameters $(r,t)$ and $(p,q)$ as 
\be
{2(t+1) \over 2r-1} ={p \over q} \,. 
\label{ourcondition}
\ee
By using the conditions $1 \le r \le q-1$, $1 \le t \le p-1$ and $p>q$, 
where $p$ and $q$ are coprime \cite{BPZ}, 
we can obtain a solution of Eq.\ (\ref{ourcondition}), 
which is $(r,t)=(2,1)$ with $(p,q) =(4,3)$. 
Therefore we obtain the identification 
$\Phi_{2,1}(z,\overline z) =i \psi(z) \overline\psi(\overline z)$ 
since $\Delta_{2,1} ={1 \over 2}$ 
in the theory with $c_{4,3} ={1 \over 2}$, 
where $\psi$ and $\overline\psi$ are 
free massless Majorana fermion fields. 
Thus, the appearance of the correlation function 
with logarithmic behaviour in the gravitational dressing of 
the Majorana fermion field theory \cite{BK} can be also understood 
in our formulation. 


\newpage



\end{document}